\documentclass[%
aip,
amsmath,amssymb,
reprint,%
]{revtex4-2}
\usepackage{graphicx}
\usepackage{dcolumn}
\usepackage{bm}

\usepackage[utf8]{inputenc}
\usepackage[T1]{fontenc}
\usepackage{mathptmx}
\usepackage{etoolbox}

\usepackage{hyperref}
\usepackage{textcomp}
\hypersetup{
	colorlinks,
	citecolor=blue,
	filecolor=black,
	linkcolor=blue,
	urlcolor=black
}
\usepackage{chemformula} 
\usepackage[capitalise]{cleveref}
\usepackage{upgreek}

\makeatletter
\def\@email#1#2{%
	\endgroup
	\patchcmd{\titleblock@produce}
	{\frontmatter@RRAPformat}
	{\frontmatter@RRAPformat{\produce@RRAP{*#1\href{mailto:#2}{#2}}}\frontmatter@RRAPformat}
	{}{}
}%
\makeatother
\begin{document}
	
	
	\title{Measurements of absolute bandgap deformation-potentials of optically-bright bilayer WSe\textsubscript{2}}
	\author{Indrajeet Dhananjay Prasad}
	\affiliation{School of Physical Sciences, Indian Institute of Technology Goa, Ponda, 403401, Goa, India}
	
	\author{Sumitra Shit}
	\affiliation{School of Physical Sciences, Indian Institute of Technology Goa, Ponda, 403401, Goa, India}
	\author{Yunus Waheed}
	\affiliation{School of Physical Sciences, Indian Institute of Technology Goa, Ponda, 403401, Goa, India}
	\author{Jithin Thoppil Surendran}
	\affiliation{School of Physical Sciences, Indian Institute of Technology Goa, Ponda, 403401, Goa, India}
	\author{Kenji Watanabe}
	\affiliation{Research Center for Electronic and Optical Materials, National Institute for Materials Science, 1-1 Namiki, Tsukuba 305-0044, Japan}
	\author{Takashi Taniguchi}
	\affiliation{Research Center for Materials Nanoarchitectonics, National Institute for Materials Science, 1-1 Namiki, Tsukuba, 305-0044, Japan}
	\author{Santosh Kumar}
	\affiliation{School of Physical Sciences, Indian Institute of Technology Goa, Ponda, 403401, Goa, India}
	\email{skumar@iitgoa.ac.in}
	
	\date{\today}
	
	\begin{abstract}
		Bilayers of transition-metal dichalcogenides show many exciting features, including long-lived interlayer excitons and wide bandgap tunability using strain. Not many investigations on experimental determinations of deformation potentials relating changes in optoelectronic properties of bilayer WSe\textsubscript{2} with the strain are present in the literature. Our experimental study focuses on three widely investigated high-symmetry points, K\textsubscript{c}, K\textsubscript{v}, and Q\textsubscript{c}, where subscript \text{c (v)} refers to the conduction (valence) band, in the Brillouin zone of bilayer WSe\textsubscript{2}. Using local biaxial strains produced by nanoparticle stressors, a theoretical model, and by performing the spatially- and spectrally-resolved photoluminescence measurements, we determine absolute deformation potential of -5.10 $\pm$\,0.24\,eV for Q\textsubscript{c}-K\textsubscript{v} indirect bandgap and -8.50\,$\pm$\,0.92\,eV for K\textsubscript{c}-K\textsubscript{v} direct bandgap of bilayer WSe\textsubscript{2}. We also show that $\approx$0.9\% biaxial tensile strain is required to convert an indirect bandgap bilayer WSe\textsubscript{2} into a direct bandgap semiconductor. Moreover, we also show that a relatively small amount of localized strain $\approx$0.4\% is required to make a bilayer WSe\textsubscript{2} as optically bright as an unstrained monolayer WSe\textsubscript{2}. The bandgap deformation potentials measured here will drive advances in flexible electronics, sensors, and optoelectronic- and quantum photonic- devices through precise strain engineering.		
	\end{abstract}
	
	\maketitle

	Mechano-optoelectronic systems utilizing transition metal dichalcogenide (TMD) semiconductors and related layered materials offer numerous opportunities for designing the heterostructure and devices with exotic optoelectronic properties. Many of these properties also depend on the number of layers in the materials of interest. Indirect-to-direct bandgap transitions, leading to efficient light-matter interactions, are both fundamentally and technologically important features of thinning down to monolayer (ML) thicknesses of the widely investigated Mo- and W- based TMD semiconductors like MoSe\textsubscript{2}, MoS\textsubscript{2}, WSe\textsubscript{2}, and WS\textsubscript{2}.\cite{mak2010atomically,wang2018colloquium,manzeli20172d,butler2013progress} Local or global deformations\cite{kumar2015strain,branny2017deterministic,palacios2017large,iff2019strain} of layered TMD semiconductors provide additional opportunities for strain engineering the bandgaps and associated optoelectronic properties\cite{peng2020strain} of these materials. Some popular ways of creating uniaxial strain in the layered materials include but are not limited to bending/stretching the flexible substrates carrying the layered materials,\cite{carrascoso2021strain,desai2014strain,tang2020strain,schmidt2016reversible} and forming wrinkles deliberately while transferring the layered materials on the standard substrates\cite{dhakal2017local}. The other standard methods of creating biaxial strains in the layered material are the creation of artificial bubbles\cite{yang2017tuning}, presence of naturally occurring bubbles\cite{guo2020direct}, using a piezo-electric substrate\cite{iff2019strain} and transferring the layered materials on pre-patterned substrates.\cite{branny2017deterministic,palacios2017large}
	
	\par Bilayer (BL)- and thicker- WSe\textsubscript{2} have indirect bandgap at Q\textsubscript{c}-K\textsubscript{v} points in the Brillouin zone (BZ), where subscript \text{c (v)} refers to the conduction (valence) band. Desai et al. \cite{desai2014strain} have shown that by creating only a moderate uniaxial tensile strain of up to $\approx$1.5\% in the BL-WSe\textsubscript{2}, it can be converted into an efficient light-emitting material. Thus, for strain engineering the optoelectronic properties of layered TMD and related materials, it is imperative to have understanding of the bands deformation potentials (DPs), which describe the energetic changes in the conduction (valence) band minima (maxima) per unit strain at the various high-symmetry points in the BZ of the materials. The importance of DP has been highlighted by Bardeen et al., see Ref.\,[\onlinecite{bardeen1950deformation}], that it provides a quantitative measure of the electron-phonon scattering phenomenon and hence explains various electrical transport properties.\cite{jin2023high,bardeen1950deformation} The absolute bandgap DPs of uniaxially-strained MLs of various layered materials have been calculated using \emph{ab initio} calculation methods.\cite{wiktor2016absolute,johari2012tuning}. Wiktor et al., see Ref.\,[\onlinecite{wiktor2016absolute}], have calculated the absolute bandgap DPs for ML-WSe\textsubscript{2} using two different approximations and have obtained its values in the range of -4.92\,to\,-5.86\,eV, and a recent experimental observation has also obtained a similar value of -6.3\,$\pm$\,0.5\,eV.\cite{sun2019stress} The reports on such theoretical calculations and experimental determinations of these bandgap DPs for BL-WSe\textsubscript{2} are missing in the literature. BL-WSe\textsubscript{2} is gaining significant attention from researchers and technologists because of its unique properties of formation of long-lived interlayer excitons with intriguing spin-valley dynamics.\cite{zhu2014anomalously,scuri2020electrically,li2017layer,jiang2021interlayer} The measurement of absolute DP of bandgap at various high-symmetry points in BZ of BL-WSe\textsubscript{2}, along with precise strain engineering, will drive advances in development of strain sensors, flexible devices, optoelectronic and quantum photonic devices.
	
	\begin{figure*}[t]
		\includegraphics[width=17cm]{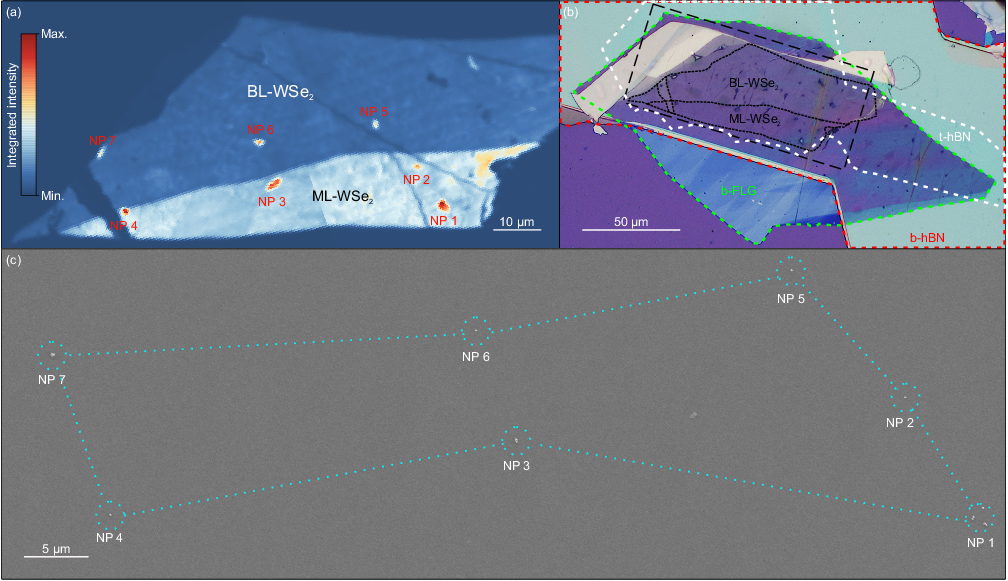}
		\caption{\textbf{NPs induced enhancement of PL signal in ML- and BL-WSe\textsubscript{2}.} (a) Room temperature \textmu-PL spacemap of integrated intensity in the spectral range of 670\,-\,870\,nm. (b) The optical image of the heterostructure. The outlines of different flakes are shown with dotted lines of different shapes. A rectangular region marks the PL spacemap region shown in (a). (c) SEM image of the NPs on h-BN/FLG/SiO\textsubscript{2}/Si-substrate. The dotted lines and circles are guides to the eye, showing the presence of NPs.}
		\label{Fig:fig1}
	\end{figure*}
	
	\par Here, we experimentally determine the absolute DP ($a^\text{gap}$) of Q\textsubscript{c}-K\textsubscript{v} and K\textsubscript{c}-K\textsubscript{v}  bandgaps of BL-WSe\textsubscript{2} by performing micro-photoluminescence (\textmu-PL) spectroscopy measurements on locally deformed BL-WSe\textsubscript{2}. Such local deformations are achieved using a simple and cost-effective technique of creating biaxial tensile strain in the exfoliated flake of the WSe\textsubscript{2} layer. \cite{surendran2024nanoparticle} Biaxial tensile strain in the exfoliated flake of WSe\textsubscript{2}, containing both ML and BL regions, is created by placing it on a h-BN/FLG/SiO\textsubscript{2}/Si-substrate which is pre-spin-coated with SiO\textsubscript{2} nanoparticles (NPs). The PL emission of ML-WSe\textsubscript{2} is mainly dominated by the neutral excitonic ($X^0$) peak due to the K\textsubscript{c}-K\textsubscript{v} transition, whereas BL-WSe\textsubscript{2} shows two excitonic peaks $X^\text{0}_\text{KK}$ and $X^\text{0}_\text{QK}$ due to K\textsubscript{c}-K\textsubscript{v} and Q\textsubscript{c}-K\textsubscript{v} transitions, respectively. We investigate strain-induced changes in all these three PL emission peaks originating from both the ML and BL regions of the fully h-BN encapsulated WSe\textsubscript{2}-flake. Using a simple theoretical approach, we quantify the absolute bandgap DP of K\textsubscript{c}-K\textsubscript{v} bandgap ($a^\text{gap}_\text{KK}$) and absolute bandgap DP of Q\textsubscript{c}-K\textsubscript{v} bandgap ($a^\text{gap}_\text{QK}$). We obtain values of -8.50\,$\pm$\,0.92\,eV for $a^\text{gap}_\text{KK}$ and -5.10\,$\pm$\,0.24\,eV for $a^\text{gap}_\text{QK}$, and show that only $\approx$0.9\% biaxial tensile strain, which is also within the range of reported values (0.4\% to 2\%) in the literature\cite{desai2014strain,tang2020strain,uddin2022engineering}, is required to convert an indirect bandgap BL-WSe\textsubscript{2} into a direct bandgap semiconductor. Interestingly, only a relatively small amount of localized strain $\approx$0.4\% is required to make a BL-WSe\textsubscript{2} as optically bright as an unstrained ML-WSe\textsubscript{2}.
	\begin{figure*}[t]
		\includegraphics[width=17cm]{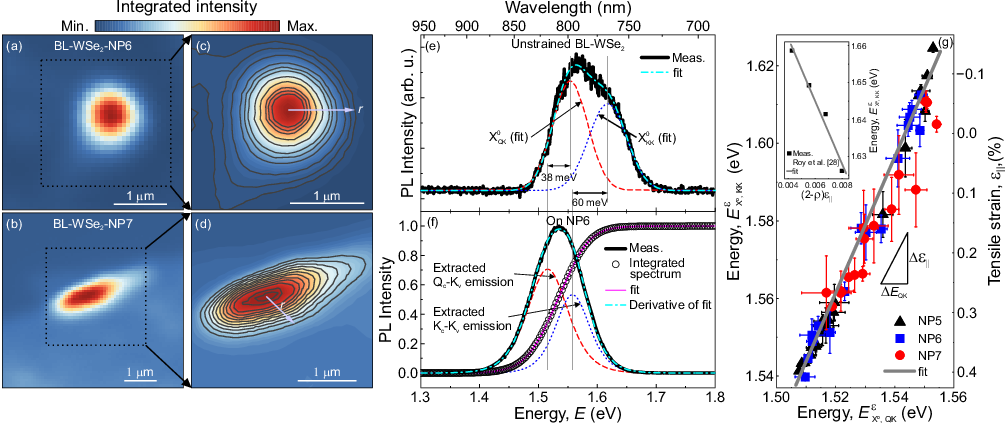}
		\caption{\textbf{\textmu-PL investigation of BL-WSe\textsubscript{2} demonstrating redshifts of both $X^\text{0}_\text{KK}$ and $X^\text{0}_\text{QK}$ emission energies under biaxial tensile strain.} \textmu-PL spacemap of integrated intensity in the spectral region of 708\,-\,918\,nm at (a) NP6 and (b) NP7 locations. (c, d) Contour maps of PL integrated intensities for the selected region (shown by dotted rectangle) in spacemaps (a) and (b), respectively. (e) A representative \textmu-PL spectrum of unstrained/flat region of BL-WSe\textsubscript{2} (thick solid line) together with two-peak Gaussian fit (dash-dotted curve) showing direct $X^\text{0}_\text{KK}$ emission peak (dotted curve) and indirect $X^\text{0}_\text{QK}$ emission peak (dashed curve). (f) \textmu-PL spectrum of strained BL-WSe\textsubscript{2} at the NP6 location (solid thick line), together with integrated intensities (open circles), fitted (solid line) using Dose-Response function with two doses representing $X^\text{0}_\text{KK}$ and $X^\text{0}_\text{KK}$ emission energies. (g) Plot of $E_{X^\text{0},\,\text{KK}}$ versus $E_{X^\text{0},\,\text{QK}}$ for various contours with varying strains at NP5 (triangle), NP6 (square), and NP7 (circle) locations showing redshifts of both $X^\text{0}_\text{KK}$ and $X^\text{0}_\text{QK}$ emission energies under biaxial tensile strain, $\epsilon_{||}$. Inset: Plot of $X^\text{0}_\text{KK}$ emission energies of ML-WSe\textsubscript{2} as a function of (2-$\rho$)$\epsilon_{||}$, where $\rho$(=0.19) is the Poisson's ratio, determining the deformation potential of K\textsubscript{c}-K\textsubscript{v} bandgap, $a^\text{gap}_\text{KK}$. Here, the emission energies and corresponding strain values were extracted from the work of Roy et al., see Ref\,[\onlinecite{roy2024upconversion}].}
		\label{Fig:fig2}
	\end{figure*}
	
	\par First, we present a theoretical model/description of absolute/hydrostatic DP, which applies both to uniaxial and biaxial strain conditions and for both ML- and BL-WSe\textsubscript{2}. The excited electrons and holes in TMD semiconductors undergo strong Coulomb interactions due to weak dielectric screening and strong geometrical confinement, forming $X^0$ and other high-order charged- and multi-excitons. Due to their large binding energies in the range of 0.2-1.0\,eV, these excitonic features are stable even at room temperature (RT), and therefore, it opens up an emerging field of building the exciton-based devices. Negligible to no changes in the $X^0$ binding energy of ML- and BL-WSe\textsubscript{2} under strain up to 3\% have been observed \cite{conley2013bandgap,shi2013quasiparticle,feng2012strain,tang2020strain,aslan2018strain}. Under these assumptions, which are also valid for our experimental conditions utilizing maximum biaxial tensile strain of only $\approx$0.4\%, the measured changes in the $X^0$ emission energy $\left(\Delta E^{\epsilon}_{X^{0}}\right)$ would be approximately equal to the changes in the bandgap energy $\left(\Delta E_\text{gap}\right)$. Thus, the absolute/hydrostatic bandgap DP can be defined as\cite{wiktor2016absolute,kumar2014anomalous,janotti2007absolute,chuang2012physics}
	\begin{equation}\label{eq:Deformation potential1}
		a^\text{gap}=\frac{\Delta E^{\epsilon}_{X^{0}}}{(\epsilon_{xx}+\epsilon_{yy}+\epsilon_{zz})}
	\end{equation}
	where $\Delta E^{\epsilon}_{X^{0}}$\,=\,$\left(E^{\epsilon}_{X^{0}} - E^{0}_{X^{0}}\right)$, $E^{\epsilon}_{X^{0}}\left(E^{0}_{X^{0}}\right)$ is the $X^0$ emission energy of strained (unstrained) WSe\textsubscript{2}, and $\epsilon_{xx}$, $\epsilon_{yy}$, and $\epsilon_{zz}$ are the normal strains in $x$, $y$, and $z$ directions, respectively. See Ref.\,[\onlinecite{wiktor2016absolute}] for a slightly different definition of the DP. Under uniaxial strain condition i.e., $\epsilon_{xx}\,=\,\epsilon_\text{axial}$, and $\epsilon_{yy}\,=\,\epsilon_{zz}\,=\,-\uprho\epsilon_\text{axial}$, where $\uprho$ is the Poisson's ratio ($\uprho$), the absolute bandgap DP can be obtained by:
	\begin{equation}\label{eq:Deformation potential axial}
		a^\text{gap}=\frac{\Delta E^{\epsilon}_{X^{0}}}{\left(1-2\uprho\right)\epsilon_\text{axial}}
	\end{equation}
	For a biaxial strain, $\epsilon_{xx}=\epsilon_{yy}=\epsilon\textsubscript{$\parallel$}$ and $\epsilon_{zz}=-\uprho \epsilon\textsubscript{$\parallel$}$,   the absolute bandgap DP can be obtained by:
	\begin{equation}\label{eq:Deformation potential biaxial}
		a^\text{gap}=\frac{\Delta E^{\epsilon}_{X^{0}}}{(2-\uprho)\epsilon\textsubscript{$\parallel$}}
	\end{equation}
	
	\section*{Results and discussion}
	\par As per definition of $a^{\text{gap}}$ in Eq.\,(\ref{eq:Deformation potential1}), its value is independent of types of strain, whether uniaxial or biaxial. We check this fact using theoretically calculated results of Desai et al., see Ref.\,[\onlinecite{desai2014strain}] for K\textsubscript{c}-K\textsubscript{v} band transition, where a redshift $\Delta E=\text{-57\,meV}$ has been calculated for 1\% uniaxial tensile strain. Using Eq.\,(\ref{eq:Deformation potential axial}), which is for uniaxial strain, we obtain DP, $a^{\text{gap}}_{\text{KK}}$\,=\,-9.19\,eV. Wiktor et al., see Ref.\,[\onlinecite{wiktor2016absolute}], have defined the DP slightly differently as the ratio of energetic changes per unit uniaxial strain and reported the theoretical DP values in the range -4.92\,eV to -5.86\,eV. Upon conversion using our theoretical model in Eq.\,(\ref{eq:Deformation potential axial}), these value span from -7.94\,eV\,to\,-9.45\,eV. Moreover, Sun et al., see Ref.[\onlinecite{sun2019stress}], have used a similar definition of DP and obtained the experimental DP value of -6.3\,$\pm$\,0.5\,eV, which on conversion using our model in Eq.\,(\ref{eq:Deformation potential axial}), we obtain a DP value of -10.16\,eV. These DP values are similar to the value we obtained using the biaxial tensile strain condition in ML-WSe\textsubscript{2}. These striking observations lead us to conclude that a particular gap, e.g., K\textsubscript{c}-K\textsubscript{v}, whether in ML or BL-WSe\textsubscript{2}, would show similar DPs. Based on this hypothesis, we started analyzing the PL emission from ML and BL-WSe\textsubscript{2} to determine Q\textsubscript{c}-K\textsubscript{v} bandgap DP.
	\par The \textmu-PL spacemap for integrated intensity of emission in the spectral range 670\,-\,870\,nm from a large size WSe\textsubscript{2} flake consisting of ML and BL region is shown in Fig.\,\ref{Fig:fig1}(a). As our scanners cover a maximum area of 50\,\textmu m$\times$50\,\textmu m at RT, two \textmu-PL spacemaps of the active layer were taken to cover ML and BL regions of the WSe\textsubscript{2} flake. The optical micrograph of the sample, marking the region of the flake from where the PL spacemap is taken, is shown in Fig.\,\ref{Fig:fig1}(b). The investigated heterostructure consists of various layers, marked with different shapes in optical micrograph as shown in Fig.\,\ref{Fig:fig1}(b). This heterostructure contains NPs underneath the WSe\textsubscript{2} layer. The \textmu-PL spacemap (Fig.\,\ref{Fig:fig1}(a)) displays a few points at which the PL signal has enhanced. Looking at the SEM image in Fig.\,\ref{Fig:fig1}(c), it is evident that the presence of these seven NPs in both ML- and BL-WSe\textsubscript{2} have created PL hotspots shown in Fig.\,\ref{Fig:fig1}(a). The local strains created by NPs are responsible for enhancing the PL signals owing to the strain-induced funneling effect.\cite{branny2017deterministic,moon2020dynamic,palacios2017large} As opposed to previous observations where BL-WSe\textsubscript{2} is considered as optically dark in PL emission, here we can see that the slightly strained (0.39\% tensile strain, quantified later) BL-WSe\textsubscript{2} is as bright as the unstrained ML-WSe\textsubscript{2} (see PL spectra in Fig.\,\ref{Fig:fig4}(c)).
	\begin{figure}[t]
		\includegraphics[width=8.5cm]{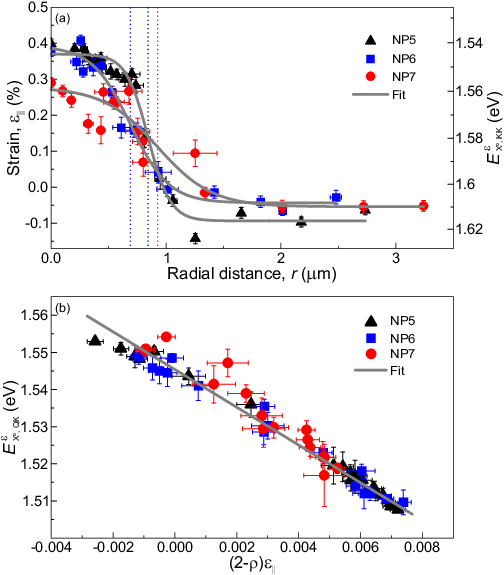}
		\caption{\textbf{Determination of absolute deformation potential (DP) of Q\textsubscript{c}-K\textsubscript{v} bandgap in BL-WSe\textsubscript{2}.} (a) The local strain, $\epsilon_{||}$, estimated using $X^\text{0}_\text{KK}$ emisssion enegies and $a^\text{gap}_\text{KK}$, is plotted as a function of contour radius, $r$ for all NPs in BL-WSe\textsubscript{2}. Solid lines are the fits of the Boltzmann sigmoidal function. The vertical dotted lines are indicating the values of characteristic radii, $r_0$, quantifying conformality of the flake on NPs. (b) Plot of  $X^\text{0}_\text{QK}$ emisssion enegies versus (2-$\uprho$)$\epsilon_{||}$, for obtaining the absolute DP of Q\textsubscript{c}-K\textsubscript{v} bandgap, $a^\text{gap}_\text{QK}$ in BL-WSe\textsubscript{2}. The triangles, squares, and circles are measured data for NP5, NP6, and NP7, respectively.}
		\label{Fig:fig3}
	\end{figure}
	\begin{figure*}[t]
		\includegraphics[width=17cm]{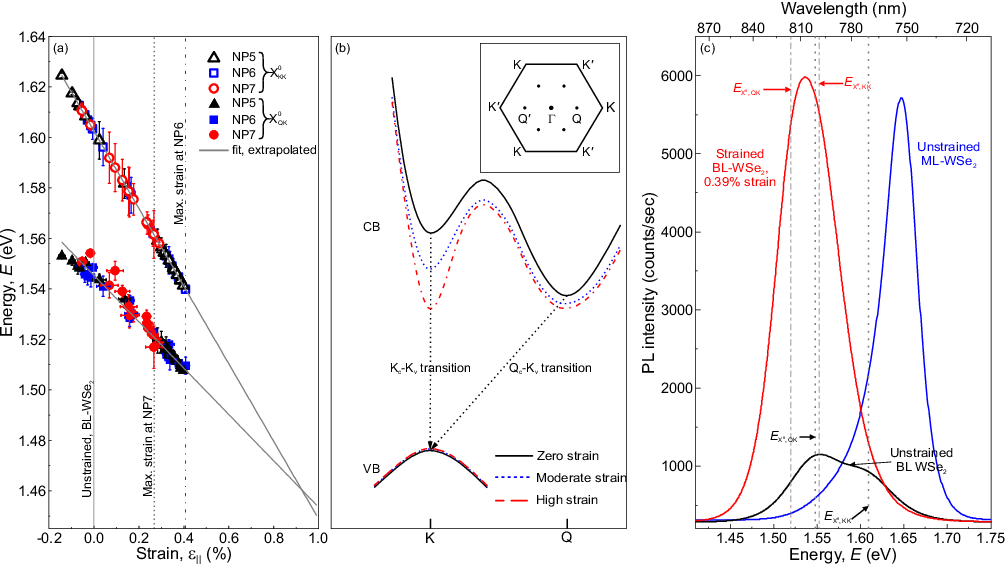}
		\caption{\textbf{Strain-induced band crossover in optically-bright BL-WSe\textsubscript{2}.} (a) Strain dependence of $X^\text{0}_\text{KK}$ nad $X^\text{0}_\text{QK}$ emission energies in BL-WSe\textsubscript{2}. Symbols are the measured energies, and solid lines are the linear fits. The vertical solid, dotted, and dash-dotted lines show zero strain, maximum strain at NP7, and maximum strain at NP6, respectively. The slopes of the fit give the gauge factors indicating that a crossover of these two energies would happen at $\approx$0.9\% strain. (b) Schematic of energy band diagram for BL-WSe\textsubscript{2}. Inset: First Brillouin zone of BL-WSe\textsubscript{2} showing high symmetry points. (c) PL emission from ML and BL-WSe\textsubscript{2} showing approximately equal signal brightness coming from $\approx$0.4\% strained BL-WSe\textsubscript{2} and unstrained ML-WSe\textsubscript{2}. The vertical dash-dotted and dotted lines provides $X^\text{0}_\text{QK}$ and $X^\text{0}_\text{KK}$ emission energies, respectively, for both unstrained and strained BL-WSe\textsubscript{2}.}
		\label{Fig:fig4}
	\end{figure*}
	\par PL emission of BL-WSe$_2$ shows excitonic emission peaks due to both indirect Q\textsubscript{c}-K\textsubscript{v} and direct K\textsubscript{c}-K\textsubscript{v} transitions \cite{aslan2020strained,lindlau2018role,desai2014strain,tang2020strain}. We looked into the details of PL signals originating from the BL-WSe\textsubscript{2} by performing high-resolution spacemaps (step size of 83\,nm). The integrated intensity \textmu-PL spacemaps around NP6 and NP7 are shown in Fig.\,\ref{Fig:fig2}(a) and Fig.\,\ref{Fig:fig2}(b), respectively. The corresponding contour spacemaps of these \textmu-PL spacemaps are shown in Fig.\,\ref{Fig:fig2}(c) and Fig.\,\ref{Fig:fig2}(d), respectively. As shown in Fig.\,\ref{Fig:fig2}(e), the BL spectra consist of two peaks attributed to direct and indirect peaks.\cite{aslan2020strained, desai2014strain,tang2020strain} A large sample of PL spectra over the entire BL-WSe\textsubscript{2} region (omitting spectra near the NPs within 3 \textmu m radii) were fitted with two-peak Gaussian function (a sample fit is shown in Fig.\,\ref{Fig:fig2}(e)), and statistical average of energies of both peaks were taken to obtain the emission energy of direct peak of K\textsubscript{c}-K\textsubscript{v} transition ($E^\text{0}_{X^\text{0},\,\text{KK}}$) as 1603\,$\pm$\,8\,meV and indirect peak of Q\textsubscript{c}-K\textsubscript{v} transition ($E^\text{0}_{X^\text{0},\,\text{QK}}$) as 1548\,$\pm$\,5\,meV for unstrained BL-WSe\textsubscript{2}. We measure the energetic difference in $X^\text{0}_\text{KK}$ and $X^\text{0}_\text{QK}$ emission energies to be 55$\pm$\,9\,meV, which lies within the reported range of 47\,to\,100\,meV in the unstrained/flat region of BL-WSe\textsubscript{2}.\cite{desai2014strain,tang2020strain,aslan2020strained,uddin2022engineering}
	
	\par As per previous reports\cite{desai2014strain, tang2020strain,uddin2022engineering}, both the direct and indirect emission peaks of BL-WSe$_2$ come energetically close to each other at higher tensile strains. This leads to the difficulty of picking these two peaks using simple two-peak Gaussian function fitting. To pick up these two peaks even at higher strains, we utilized the following method consisting of three steps; firstly normalized PL spectra between 0 and 1, secondly integrated this normalized data over the whole spectrum range, and lastly, fitted this integrated-normalized data using Dose-Response function with two doses representing $E^\epsilon_{X^\text{0},\,\text{QK}}$ and $E^\epsilon_{X^\text{0},\,\text{KK}}$ values (See Supporting Information for details). Figure\,\ref{Fig:fig2}(f) shows an example of such fitting for PL spectrum taken at the NP6 location in BL-WSe\textsubscript{2}. The open circles are the integrated intensity of normalized measured data. The solid curve on these open circles shows the fitted data. The derivative of this fitted curve is shown with a dash-dotted curve on top of normalized measurement data. The fitted parameters give us both $X^\text{0}_\text{QK}$ and $X^\text{0}_\text{KK}$ emission energies indicated by vertical lines in Fig.\,\ref{Fig:fig2}(f). We de-convolved and differentiated the integrated intensities fitted data to obtain the emission peaks of both these transitions, which are shown in Fig.\,\ref{Fig:fig2}(f) with the dashed and dotted curves for $X^\text{0}_\text{QK}$ and $X^\text{0}_\text{KK}$ emissions, respectively. It can be observed that in comparison to unstrained region (see Fig.\,\ref{Fig:fig2}(e)), emission energies of both excitons, $X^\text{0}_\text{QK}$ and $X^\text{0}_\text{KK}$, are redshifted in the strained region (see Fig.\,\ref{Fig:fig2}(f)) by 38\,meV and 60\,meV, respectively. As can be seen from both the contour maps in Fig.\,\ref{Fig:fig2}(c,\,d), when going away from the center of the hotspot, the integrated intensity is decreasing significantly. This variation in the intensity indicates that the local biaxial tensile strain is decreasing with an increase in radial distance. In the following, we quantify the variation of this local biaxial tensile strain as a function of radial distance.

	\par The biaxial tensile strain for different contours is determined by analyzing the PL spectra corresponding to pixels of the spacemap where the contour lines are intersecting. For every point on the contour, we obtain the distance of that point from the center of the contour, and by fitting the PL spectra collected at that point using the Dose-Response function, as mentioned above, we extract $X^\text{0}_\text{QK}$ and $X^\text{0}_\text{KK}$ emission energies. As each contour consists of many such points, the statistical average of all these distances gives us a radial distance ($r$) of a particular contour; the statistical average of the $X^\text{0}_\text{QK}$ and $X^\text{0}_\text{KK}$ emission energies from all these points gives us the $E^\epsilon_{X^\text{0},\,\text{QK}}$ and $E^\epsilon_{X^\text{0},\,\text{KK}}$ values, respectively for that contour with local strain $\epsilon$. The error bars have also been obtained from the statistical analysis for $E^\epsilon_{X^\text{0},\,\text{QK}}$, $E^\epsilon_{X^\text{0},\,\text{KK}}$ and $r$ values. In contrary to the previous reports, which say that $X^\text{0}_\text{QK}$ emission energy blueshifts as a function of tensile strain, here our work, as presented in Fig.\,\ref{Fig:fig2}(g), clearly shows that both $E^\epsilon_{X^\text{0},\,\text{QK}}$ and $E^\epsilon_{X^\text{0},\,\text{KK}}$ redshift as we move from unstrained (flat) region to the tensile strained (towards center of the hotspot) region. It can also be observed from Fig.\,\ref{Fig:fig2}(g) that the rate of change of $X$\textsuperscript{0} emission energy per percent biaxial tensile strain for K\textsubscript{c}-K\textsubscript{v} bandgap is larger than the Q\textsubscript{c}-K\textsubscript{v} bandgap. In the following, we understand these unequal energy changes through the determination of absolute bandgap DPs of K\textsubscript{c}-K\textsubscript{v} and Q\textsubscript{c}-K\textsubscript{v} bandgaps.
	
	\par Extensive work has been done in the literature to find the neutral excitonic energy-gauge factor of K\textsubscript{c}-K\textsubscript{v} transition in ML-WSe\textsubscript{2}, and its values in the range of 41 to 153\,meV/\% strain have been reported.\cite{schmidt2016reversible,desai2014strain,carrascoso2021strain,roy2024upconversion,frisenda2017biaxial} We have followed the work of Roy et al., see Ref.\,[\onlinecite{roy2024upconversion}], where authors have performed both Raman- and PL-spectroscopy measurements. From their work, we extracted values of measured $X^\text{0}$ emission energies at K\textsubscript{c}-K\textsubscript{v} bandgap and biaxial tensile strain. We plotted those energies as a function of (2-$\rho$)$\epsilon_{||}$, where $\rho$\,=\,0.19\,\cite{zeng2015electronic,kang2013band,ccakir2014mechanical,falin2021mechanical}. This plot is shown in the inset of Fig.\,\ref{Fig:fig2}(g). As per Eq.\,(\ref{eq:Deformation potential biaxial}), the slope of this plot gives value of the $a^{\text{gap}}_{\text{KK}}$ as -8.50\,$\pm$\,0.92\,eV. As discussed earlier, the DP of K\textsubscript{c}-K\textsubscript{v} bandgap shall have same value in both ML- and BL-WSe\textsubscript{2} and hence, $a^{\text{gap}}_{\text{KK}}$ along with Eq.\,(\ref{eq:Deformation potential biaxial}) was utilized for estimating the biaxial strain in BL-WSe\textsubscript{2}. The biaxial strain ($\epsilon_{||}$) is plotted in Fig.\,\ref{Fig:fig3}(a) as a function of $r$. The solid lines are the Boltzmann sigmoidal function fits. From the fit, we obtain the characteristic parameter, $r_0$, which provides the radial distance at which the strain becomes half the maximum strain. The diameter of conformality (2$r_0$) for NP5, NP6, and NP7 are 0.84\,\textmu m, 0.69\,\textmu m, and 0.93\,\textmu m, respectively, which are larger than the measured spatial resolution of the setup (see methods). This explains lower strain values at these points than a theoretical prediction of 0.92\% (see Supporting Information). The variations of local strains for some NPs are stronger than others due to different levels of conformality of the flakes at the NP locations. NP5 and NP6 show similar maximum strain of $\approx$0.40\%, but NP7 shows a smaller strain of 0.29\% because it is present at the edge of the flake. Now, we plot $E^\epsilon_{X^\text{0},\,\text{QK}}$ as a function of (2-$\uprho$)$\epsilon_{||}$, as shown in Fig.\,\ref{Fig:fig3}(b). As per Eq.\,(\ref{eq:Deformation potential biaxial}), the slope of $E^\epsilon_{X^\text{0},\,\text{QK}}$ vs (2-$\uprho$)$\epsilon_{||}$ gives the value of $a^{\text{gap}}_{\text{QK}}$, which is -5.10 $\pm$\,0.24\,eV. The gauge factor ($\Delta E_{\text{QK}}/\Delta \epsilon_{||}$), a widely used physical parameter in optomechanical engineering and material sciences can be estimated either from Fig.\,\ref{Fig:fig3}(b) or Fig.\,\ref{Fig:fig2}(g), which is equal to -102.6\,$\pm$\,2.5\,meV/\% biaxial tensile strain.
	
	\setcounter{figure}{0}
	\renewcommand{\thefigure}{S\arabic{figure}}
	\begin{figure*}
		\includegraphics[width=17cm]{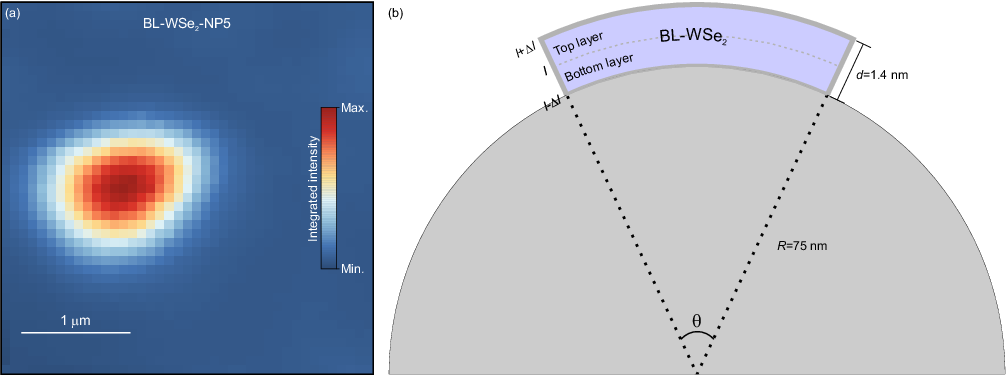}
		\caption{\textbf{\textmu -PL spacemap and strain estimation using bending beam theory.} (a) \textmu -PL spacemap around NP5 in BL-WSe\textsubscript{2}. (b) Upper half of a NP is shown by hemisphere and BL-WSe\textsubscript{2} is placed on top of the hemisphere. }
		\label{Fig:fig_s1}
	\end{figure*}
	
	\par Lastly, we plot both $E^\epsilon_{X^\text{0},\,\text{KK}}$ and $E^\epsilon_{X^\text{0},\,\text{QK}}$ as a function of $\epsilon_{||}$ in Fig.\,\ref{Fig:fig4}(a) and we see that $E^\epsilon_{X^\text{0},\,\text{KK}}$ is decreasing faster than $E^\epsilon_{X^\text{0},\,\text{QK}}$ with an increase in biaxial tensile strain. It has been shown by Shen et al., see Ref.\,[\onlinecite{shen2016strain}], that the valance bands of WSe\textsubscript{2} are not much affected by strain up to 2\%, and thus we ascribe these reductions in $X^\text{0}_\text{KK}$ and $X^\text{0}_\text{QK}$ emission energies with biaxial tensile strain due to decrease in the conduction band minima at K and Q points, respectively as shown schematically in Fig.\,\ref{Fig:fig4}(b). It also shows the expected changes in the conduction band edges for unstrained (solid curve), moderate (dotted curve), and high strain (dash-dotted curve) conditions. As the measured absolute DP of K\textsubscript{c}-K\textsubscript{v} bandgap is larger than that of Q\textsubscript{c}-K\textsubscript{v} bandgap, when we increase the tensile strain, $X^\text{0}_\text{KK}$ emission energy would decrease at a higher rate than $X^\text{0}_\text{QK}$ emission energy as shown by solid fitted lines in Fig.\,\ref{Fig:fig4}(a). This indicates that BL-WSe\textsubscript{2} will undergo a transition from an indirect bandgap to a direct bandgap semiconductor at merely 0.9\% biaxial tensile strain. Using Eqs.\,(\ref{eq:Deformation potential axial})\,and\,(\ref{eq:Deformation potential biaxial}), we estimate that $\approx$2.63\% uniaxial tensile strain is required to achieve the same effect. From PL spectra of ML and BL-WSe\textsubscript{2} in Fig.\,\ref{Fig:fig4}(c), it is striking to observe that in our work, we have achieved only $\approx$0.4\% of maximum biaxial tensile strain, and even at this low level of strain, the PL emission from BL-WSe\textsubscript{2} is equally brighter as the ML-WSe\textsubscript{2} emission. As electron-phonon scattering\cite{bardeen1950deformation,jin2023high} is proportional to $\text{DP}^{2}/\text{Y}$, where Y is Young's modulus of elasticity, we see clearly that electrons in BL-WSe\textsubscript{2} are less scattered with longitudinal acoustic phonons than in ML-WSe\textsubscript{2}, which may explain the brightness of BL-WSe\textsubscript{2}.
	\par In summary, the deformations produced by uniaxial or biaxial tensile strain provide opportunities for strain engineering the bandgap and associated optoelectronic properties. The quantification of energetic changes in conduction band minima per unit strain at various high symmetry points in the BZ, expressed as a parameter named absolute bandgap DP, leads to the prediction of electron-phonon scattering rate apart from its applications in strain-based devices; which in turn provides information on electrical transport properties. Numerous efforts have been made to determine the bandgap  DP of various bands in ML-WSe\textsubscript{2}, but for the BL-WSe\textsubscript{2}, which possesses unique properties such as the formation of interlayer excitons with intriguing spin-valley dynamics, theoretical and experimental values of bandgap DP are missing in the literature. We experimentally determined the bandgap DP of the indirect and direct bandgap in BL-WSe\textsubscript{2} by inducing biaxial tensile strain in BL-WSe\textsubscript{2} using nanoparticle stressors. Both the reported theoretical and experimental absolute DP values, along with our theoretical model and experimental findings indicate that the K\textsubscript{c}-K\textsubscript{v} bandgap DP should be independent of the strain type and whether material is ML- or  BL-WSe\textsubscript{2}. We determined the absolute DP of the K\textsubscript{c}-K\textsubscript{v} bandgap in ML-WSe\textsubscript{2} to be -8.50\,$\pm$\,0.92\,eV and used this value for the K\textsubscript{c}-K\textsubscript{v} bandgap in BL-WSe\textsubscript{2} to predict the biaxial tensile strain induced by NPs. By analyzing the exciton emission from the Q\textsubscript{c}-K\textsubscript{v} bandgap at different strain values, we obtained the absolute DP of the Q\textsubscript{c}-K\textsubscript{v} bandgap equal to -5.10\,$\pm$\,0.24\,eV. Furthermore, we determined the gauge factor for exciton emission from Q\textsubscript{c}-K\textsubscript{v} bandgap in BL-WSe\textsubscript{2} as -102.6\,$\pm$\,2.5\,meV/\% biaxial tensile strain. Interestingly, light emission from $\approx$0.4\% strained, and an indirect bandgap BL-WSe\textsubscript{2} material has a similar intensity of unstrained direct bandgap ML-WSe\textsubscript{2} material; additionally, we predict that BL-WSe\textsubscript{2} will undergo a transition from indirect bandgap to direct bandgap semiconductor at 0.9\% biaxial tensile strain whereas 2.63\% uniaxial tensile strain is required to achieve the same effect. We have used mechanical exfoliation techniques to extract ML and BL from the whole crystal and a dry transfer technique to transfer the flakes onto the substrate.
	\section*{Methods}
	\subsection*{Sample preparation}
	\par The bottommost layer of the investigated heterostructure is Few Layer Graphene (FLG), placed on SiO\textsubscript{2}/Si substrate. The SiO\textsubscript{2} is 285\,nm thick. The second layer on top of FLG is a thick h-BN layer. SiO\textsubscript{2} NPs were spin-coated at 8000 rpm on the h-BN layer, and the SEM image was taken to confirm the presence of the NPs on the h-BN layer. Now, an active layer of WSe\textsubscript{2} was transferred using a dry transfer technique\,\cite{castellanos2014deterministic}, and it was further encapsulated with a thin h-BN layer. The active layer of WSe\textsubscript{2} is encapsulated with h-BN from the top to protect from the environment and from the bottom to reduce substrate effects. FLG is the lowermost layer; in the future, the heterostructure will also be capped with FLG from the top to carry out electrical measurements.
	\subsection*{PL spectroscopy}
	\par PL signals from samples were collected using a home-built confocal microscope, yielding a diffraction-limited spatial resolution of 0.54\,\textmu m, which is slightly more than the theoretical diffraction limit of 0.50\,\textmu m. The samples were mounted on an XYZ nanopositioner, and an objective of 0.82 NA was used to collect signals from the sample. For generating the space maps of PL signals, the samples were scanned using an XY scanner providing a scanning range of 50\,\textmu m at RT. A diode-pumped solid-state CW laser emitting at $\lambda$ = 532\,nm was used for PL measurements. An ultrasteep longpass filter designed at an edge of 533.3\,nm was used to suppress the laser light from entering the spectrometer. All the spectra were acquired with a 0.5\,m focal length spectrometer and water-cooled charge-coupled device providing a best spectral resolution of $\approx$2.5\,meV at $\lambda$ = 532\,nm on a 150 lines/mm grating.
	
    \section*{Author contributions}
    I.D.P. carried out micro-photoluminescence measurements supported by Y.W. and J.T.S. under the supervision of S.K.. S.S. and J.T.S. fabricated the sample heterostructure under the supervision of S.K.. I.D.P. and S.K. analyzed the data, developed a theoretical model, and interpreted the experimental and theoretical modeling results. I.D.P. and S.K. wrote the manuscript. All authors discussed the results and contributed to the manuscript. S.K. conceived and coordinated the project.

 	\begin{acknowledgments}
		We thank E.S. Kannan, S. R. Parne, and A. Rahman for the fruitful discussion and A. Rastelli for the data analysis software. This work was supported by the DST Nano Mission grant (DST/NM/TUE/QM-2/2019) and the matching grant from IIT Goa. I.D.P. thanks The Council of Scientific $\&$ Industrial Research (CSIR), New Delhi, for the doctoral fellowship. K W and T T acknowledge support from the JSPS KAKENHI (Grant Numbers 21H05233 and 23H02052) and World Premier International Research Center Initiative (WPI), MEXT, Japan.
	\end{acknowledgments}
	
	\section*{Supporting information}
	\subsection*{\textmu -PL space map around nanoparticle}
	Fig.\,\ref{Fig:fig_s1}(a) shows the \textmu-PL space map around NP5 underneath BL-WSe\textsubscript{2} of the heterostructure, which is not shown in the manuscript.
	
	\subsection*{Boltzmann sigmoidal function fit}
	The Boltzmann sigmoidal function which is used to fit the strain ($\epsilon_{||}$) versus radial distance ($r$) curve shown in Fig.\,3(a):
	\begin{equation}\label{eq:Boltzmann function}\nonumber
		\epsilon = \epsilon_{min}+ \frac{\epsilon_{max} - \epsilon_{min}}{1 +\,\exp\left(\frac{r - r_0}{\delta r}\right)}
	\end{equation}	
	where $\epsilon_{max}$, $\epsilon_{min}$, and $\delta r$ represent maximum strain, minimum strain, and time constant, respectively. The characteristic parameter $r_0$ represents radial distance at which strain becomes half of its maximum value.

	\subsection*{Dose response function consisting of two doses}
	The sigmoidal dose response function was used to fit the integrated intensity curve (a sigmoidal curve, See Fig.\,2(f)) to pick up two emission peaks (corresponding to the direct and indirect excitons) for a PL spectrum from a strained BL-WSe\textsubscript{2}. The function is given as
	\begin{equation}\nonumber
		I=I_1+(I_2-I_1)\left(\frac{p}{1+10^{\left(Log(E_{1})-E\right)h_1}}+\frac{1-p}{1+10^{\left(Log(E_{2})-E\right)h_2}}\right)
	\end{equation}
	where $I\textsubscript{1}$ and $I\textsubscript{2}$ are base value and maximum value of integrated intensity, respectively. The first sigmoidal part, $\frac{p}{1+10^{\left(Log(E_{1})-E\right)h_1}}$, produces a sigmoidal tranisition around the point $E=Log(E_1)$ with steepness controlled by parameter $h_1$. Simillarly, the second sigmoidal part, $\frac{1-p}{1+10^{\left(Log(E_{2})-E\right)h_2}}$ produces a sigmoidal trasition around the point $E=Log(E_2)$ with steepness controlled by parameter $h_2$. The parameter $p$ determines the contribution of the first sigmoid, and $(1-p)$ determines the contribution of the second sigmoid in the full sigmoidal curve. The parameters $Log(E\textsubscript{1})$ and $Log(E\textsubscript{2})$ give the emission energy of the first and second peaks, respectively.
	
	\subsection*{Beam bending theory of strain estimation}
	The beam bending method was used to verify the strains that were estimated using PL signal at the NP locations. Consider a BL-WSe\textsubscript{2} material having thickness $d$\,(=\,1.4\,nm) and length $l$ for central plane. As shown in Fig.\,\ref{Fig:fig_s1}(b), when BL-WSe\textsubscript{2} is fully conformal on the NP then the bending radius of curvature of BL-WSe\textsubscript{2} would be same as the radius $R$\,(=\,75\,nm) of NP. Due to this bending as shown in Fig.\,\ref{Fig:fig_s1}(b), the length of the uppermost layer increases by $\Delta l$. From Fig.\,\ref{Fig:fig_s1}(b) geometry and using the relation between arc, radius, and angle, we can write $\theta=\frac{l}{R+\frac{d}{2}}=\frac{l+\Delta l}{R+d}$, which on simplifying gives the strain $\left(\epsilon_{||}=\frac{\Delta l}{l}=\frac{d}{2R+d}\right)$.

	\providecommand{\noopsort}[1]{}\providecommand{\singleletter}[1]{#1}%

	\end{document}